# Multi-frame Signature-cum Anomaly-based Intrusion Detection Systems (MSAIDS) to Protect Privacy of Users over Mobile Collaborative Learning (MCL)

[1]Abdul Razaque
[2]Khaled Elleithy

[1]Computer and Engineering Department
University of Bridgeport 126 Park Avenue
Bridgeport, CT 06604 USA
arazaque@bridgeport.edu

[2]Computer and Engineering Department
University of Bridgeport 126 Park Avenue
Bridgeport, CT 06604 USA
elleithy@bridgeport.edu

**Abstract.** *The rogue DHCP is unauthorized server that releases the incorrect IP address to users and sniffs the traffic illegally. The contribution specially provides privacy to users and enhances the security aspects of mobile supported collaborative framework (MSCF) explained in [24].The paper introduces multi-frame signature-cum anomaly-based intrusion detection systems (MSAIDS) supported with novel algorithms and inclusion of new rules in existing IDS. The major target of contribution is to detect the malicious attacks and blocks the illegal activities of rogue DHCP server. This innovative security mechanism reinforces the confidence of users, protects network from illicit intervention and restore the privacy of users. Finally, the paper validates the idea through simulation and compares the findings with known existing techniques.*

## 1. Introduction

The rapid advancement in network technologies has augmented the use of mobile devices in open, large scale and heterogeneous environments. The mobile devices create the base for users to exchange information anytime and anywhere in the world. The exploitation of mobile devices has not only underpinned communication but also provided many chances for malicious attackers to break the privacy of users. The mobile users are extremely reliant on DHCP server for obtaining IP address because DHCP server offers highly structured service to mobile devices.

From other side, unauthorized DHCP server (rogue DHCP) produces the problems for users and cracks the security. It invites intruders and attackers to redirect traffic of any device that uses the DHCP. Intruder modifies the original contents of communication. The malware and Trojans horse install rogue DHCP server automatically on network and affect the legitimate servers .If rogue DHCP server assigns an incorrect IP address faster than original DHCP server, it causes potentially black hole for users. To control the malicious attacks and avoiding the network blockage, the network administrators put their efforts to guarantee the components of server, using various tools. The graphical user interface (GUI) tool is used to prevent the attack of rogue detection [5]. Idea of using multilayer switches may be configured to control the attacks of rogue DHCP server but it is little bit complex and not efficient to detect rogue DHCP server.
The DHCP spoofing is another solution for detecting rogue DHCP server. However, if single segment is spoofed that can damage the whole network. Spoofing method takes long time till intruder has enough time to capture the traffic and assign wrong IP address [8]. Time-tested, DHCP Find Roadkil.net's, DHCP Sentry, Dhcploc.exe and DHCP-probe provide the solution to detect and defend rogue DHCP server malware [6]. All of these tools cannot detect the new malicious attacks [2].
Distributed Intrusion Detection System (DIDS) is another technique to support the mobile agents. This technique helps the system to sense the intrusion from incoming and outgoing traffics to detect known attacks [1]. Ant colony optimization (ACO) based distributed intrusion detection system is introduced to detect intrusions in the distributed environments. It detects the visible activities of attackers and identifies the attack of false alarm rate but it does not detect DOS attacks [3]. Anomaly based intrusion detection are introduced to detect those attacks for which no signatures exist [4], [6], [10]. This paper introduces the multi-frame signature-cum anomaly based intrusion detection system supported with novel algorithms, inclusion of new rules in existing IDS to detect malicious attacks and increase the privacy and confidentiality of users. The reminder of paper is organized as follows: The section 2 describes related work and background study. Possible attacks of rogue DHCP server are explained in Section 3. The proposed solutions including functional components are given in section 4. Simulation setup is explained in section 5. The analysis of result and discussion are given in section 6. Finally conclusion of the paper is given in section 7.

## 2. Related Work and Background Study

The modern technologies and its deployment in computer and mobile devices have not only created new opportunities for better services but from other perspective, privacy of the users is highly questionable. The network-intruder and virus contagion extremely affect the computer systems and its counterparts. They also alter the top confidential data. Handling these issues and restoring the security of systems, IDS are introduced to control malicious attackers.IDS are erroneous and not providing the persistent solution in its current shape. The first contribution in the field of intrusion detection was deliberated by J.P Anderson in [7].The author introduced notion about the security of computer systems and related threats. Initially, three attacks were discovered that are misfeasors, external penetrations and internal penetrations.
The classification of typical IDS is discussed in [17]. The focus of the contribution is about reviewing the agent-based IDS for mobile devices. They have stated problems and strength of each category of classification and suggested the methods to improve the performance of mobile agent for IDS design.

Four types of attacks are discussed in [21] for security of network. They have also simulated the behavior of these attacks by using simulation of ns2. A multi-ant colonies technique is proposed in [22] for clustering the data. It involves independent, parallel ant colonies and a queen ant agent. Authors state that each process for ant colony takes dissimilar forms of ants at moving speed. They have generated various clustering results by using ant-based clustering algorithm. The findings show that outlier's lowest strategy for choosing the recent data set has the better performance. The contribution covers the clustering-based approach.

The work done in [18] is about the framework of distributed Intrusion Detection System that supports mobile agents. The focus of work is to sense both outside and inside network division. The mobile-agents control remote sniffer, data and known attacks. The paper has introduced data mining method for detection and data analysis. Dynamic Multi-Layer Signature based (DMSIDS) is proposed in [2]. It detects looming threats by using mobile agents. Authors have introduced small and well-organized multiple databases. The small signature-based databases are also updated at the same time regularly.

In addition, all of the proposed techniques cover general idea of network detection but proposed MSAIDS technique handles the irreplaceable issues of DHCP rogue server. The contribution also prevents almost all types of DOS attacks. The major contribution of work is to validate technique by employing innovative algorithms and inclusion of new rules in existing tradition. It also helps the legitimate users to start secure and reliable communication frequently over MCL [23]. One of the most promising aspects of this research is uniqueness because there is no single contribution is available in survey about the DHCP rogue and its severe targeted attacks.

## 3. Possible Attacks of Rogue DHCP Server

The introduction of distributed system has highly affected the security [11]. There are several forms of vulnerabilities and vigorous threats to expose the security of systems. To take important security measures and enhancing the secure needs for organizations, several mechanisms are implemented but those mechanisms also invite attackers to play with privacy and confidentiality of users. One of the major threats for privacy of data is intervention of rogue DHCP server. The first sign of problem associated with rogue DHCP server is discontinuation of network service. The static and portable devices start experiencing due to network issues. The issues are started by assigning the wrong IP address to requested users to initiate the session.

The malicious attackers take the advantages of rogue DHCP server and sniff the traffic sent by legitimate users. Rogue DHCP server spreads wrong network parameters that create the bridge for intruders to expose the privacy. Trojans like DNS-changing installs the rogue DHCP server and pollutes network. Rogue DHCP server creates several problems to expose the privacy of legitimate users. We highlight two major types of security attacks to be created by rogue DHCP server.

### 3.1. Sniffing the Network Traffic

It is brutal irony in information security that the features which are used to protect static and portable devices to function in efficient and smooth manner; and from other side, same features maximize the chances for intruders to compromise and exploit the same tools and networks. Hence packet sniffing is used to monitor network traffic to prevent the network from bottleneck and make an efficient data transmission. Intruders use same resources for collecting information for illegal use. Rogue DHCP server helps malicious intruders to expose privacy of users. When networks are victim of rogue DHCP server that provides very important information related to IP address, domain name system and default gateway to intruders.

All of this information helps intruders to sniff traffic of legitimate users. Rogue DHCP server is introduced on secure environment to collect confidential information and sniffs the traffic and wreaks the privacy of users shown in figure 1.

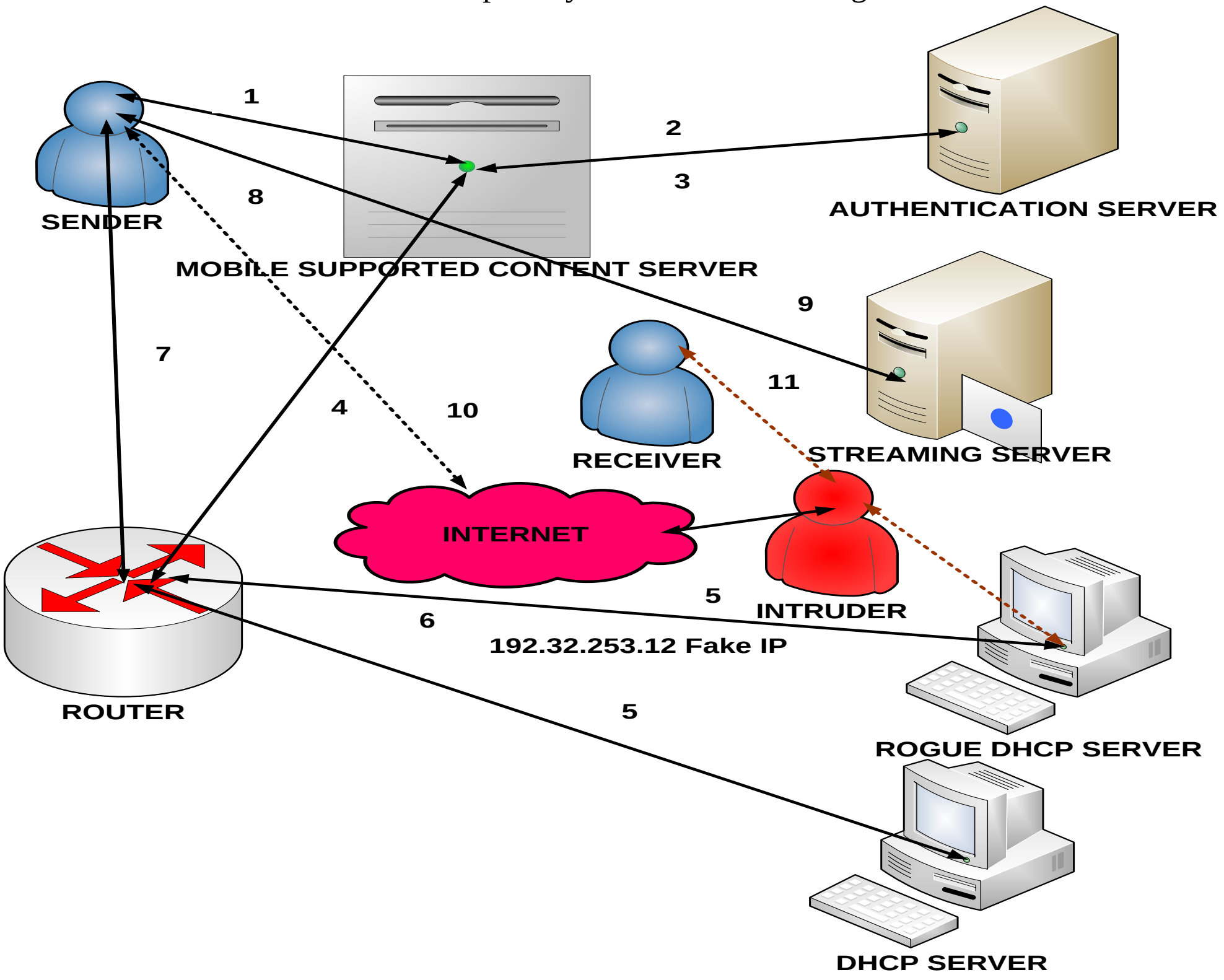

Figure 1. Sniffing the traffic and masquerading attack

Rogue DHCP server also facilitates for intruders to capture the MAC address of legitimate users. It causes sniffing the traffic through switch. In this case, intruder spoofs IP addresses of both sender and receiver and plays the man-of-middle to sniff traffic and extract important contents of communication. It causes the great attack on privacy of users.

### 3.2. Denial of Service Attack (DOS)

Intruder gets support through rogue DHCP server also uses DOS attacks after sniffing confidential contents of traffic. Due to DOS attack, the access of important services for legitimate users is blocked. Intruder often crashes routers, host, servers and other computer entity by sending overwhelming amount of traffic on the network. Rogue DHCP server creates friendly environment for intruder to launch DOS attacks because intruder needs small effort for this kind of attack and it is also difficult to detect and attack back to intruder [12]. In addition, it is also easy to create floods on internet because it is comprised of limited resources including processing power, bandwidth and storage capabilities. Rogue DHCP makes flooding attack at domain name system (DNS) because target of intruder is to prevent the legitimate users [12] & [16].
These attacks on DNS have obtained varying success while disturbing resolution of names related to targeted zone. Rogue DHCP server takes advantages of inevitable human errors during installation, configuration and developing software. It creates several types of DOS attack documented in literature [20]. Intruder with support of Rogue DHCP server makes three types of attacks: fragile (smurf), SYN Flood and DNS DOS attacks shown in figure 2. These attacks are vulnerable and dangerous for security point of view.

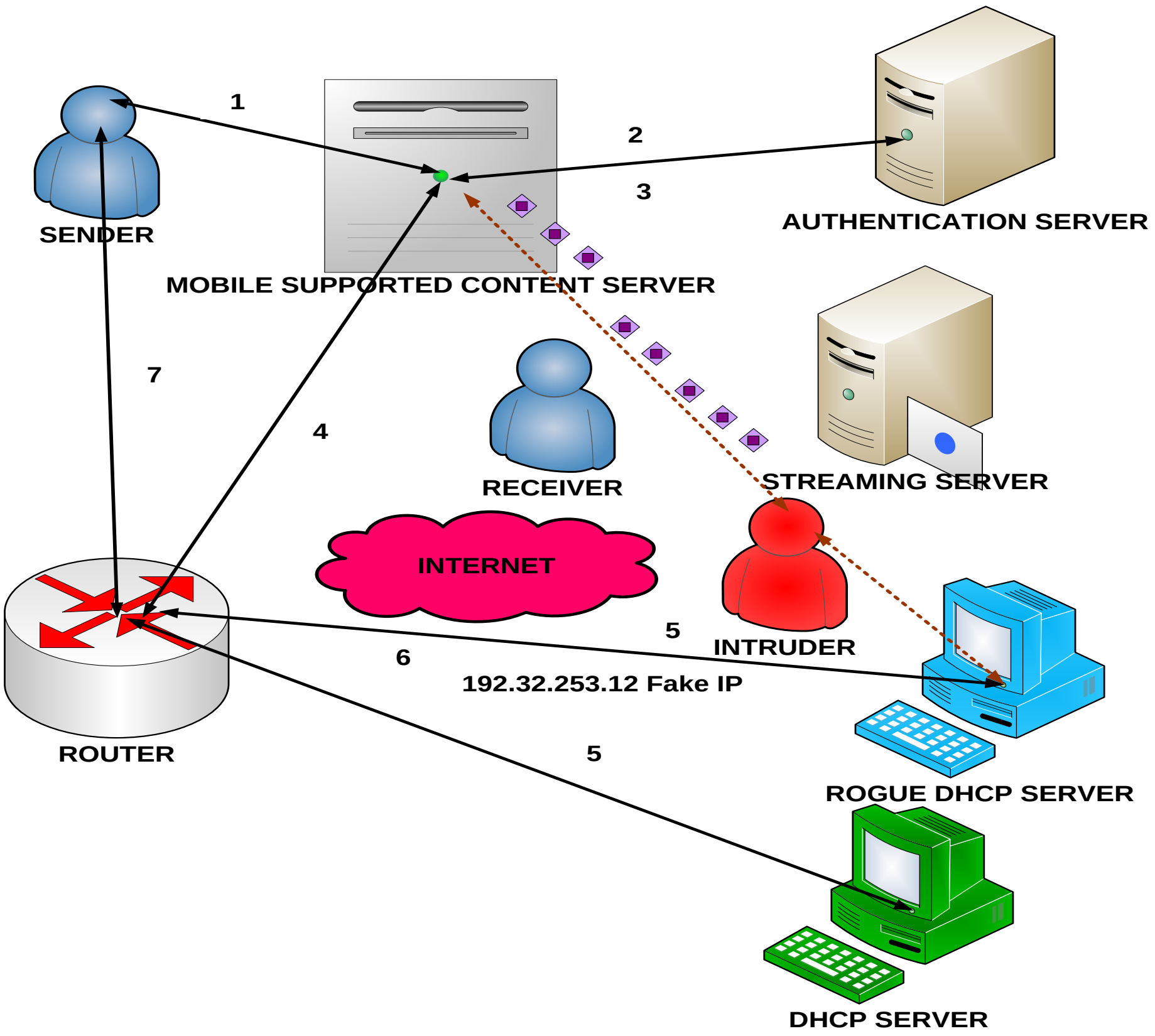

Fig.ure 2.denial of service attack (DOS) attack

## 4. Proposed Solution (Multi-frame Signature-cum-anomaly based Intrusion detection system)

Networks are being converged rapidly and thousands of heterogeneous devices are connected. The devices integrated in large networks, communicate through several types of protocols and technologies. This large scale heterogeneous environment invites the intruders to expose security of users. Hence, IDS are introduced to recognize the patterns of attacks, if they are not fixed strategically, many intruders cross IDS by traversing alternate route in network.

Many signature-based IDS are available to detect attacks but some of new attacks cannot be identified and controlled. Anomaly-based IDS is another option but it only detects limited new attacks. The multi-frame signature-cum anomaly-based intrusion detection system (MSAIDS) supported with algorithms is proposed to resolve issue of DHCP rogue. The proposed framework consists of detecting server that controls IDS and its related three units: (i) DHCP verifier unit (ii) signature database (iii) anomaly database.

During each detection process, intrusion detection starts matching from DHCP verifier, if any malicious activity is detected that stops process otherwise checks with two units until finds either malicious activity or not. Figure 3 shows MSAIDS.

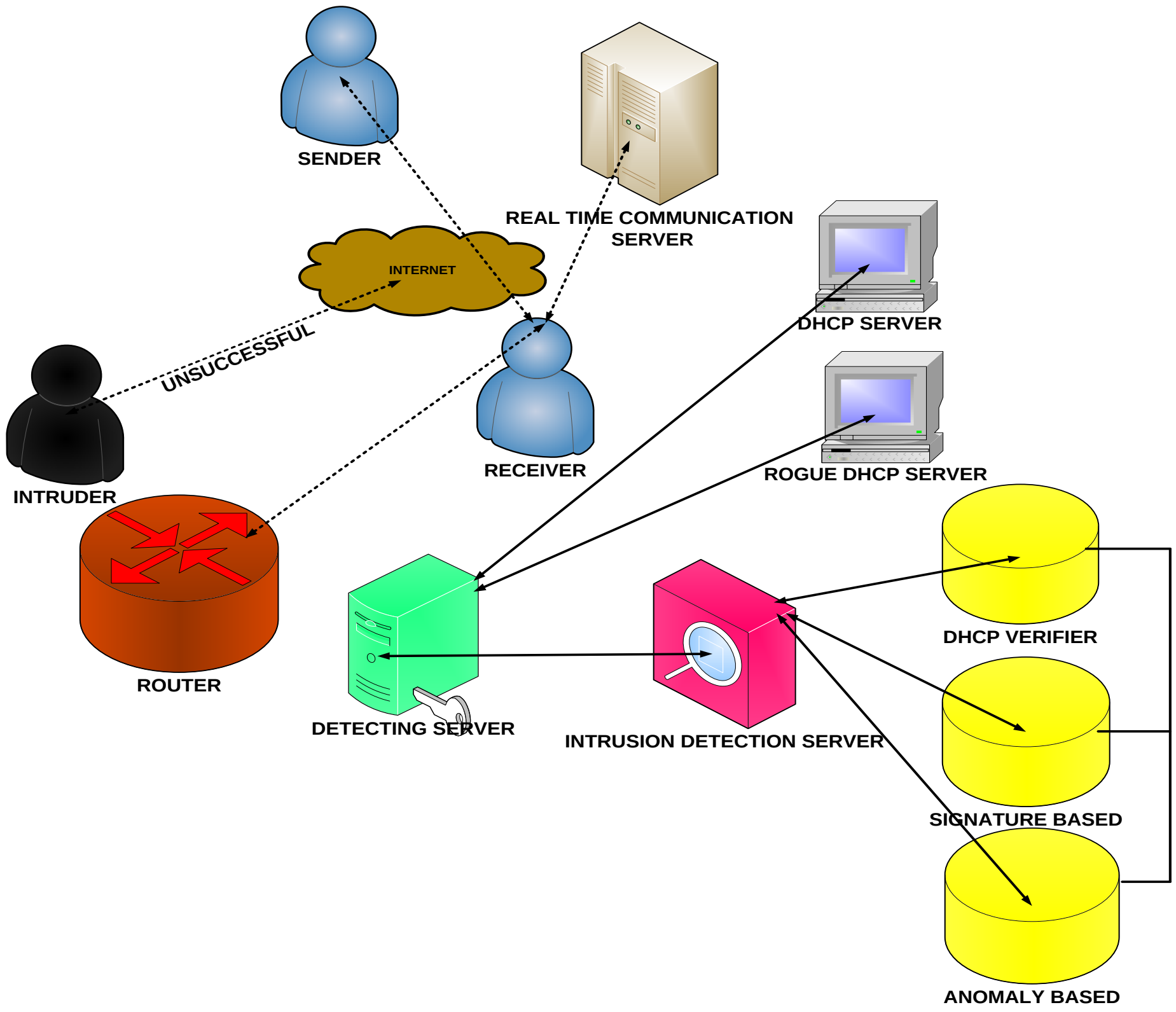

Figure 3. Multi-frame signature-cum-anomaly based IDS

The detecting server (DS) is responsible to check inbound and outbound traffic for issuance of IP address. The DS gets IP request (inbound traffic) from routers and forwards to DHCP server after satisfactory checkup. When any IP address is released for requested node then applies DHCP detecting algorithm for validation of DHCP server and detecting types of attack shown in algorithm 1.

**Algorithm 1: Verify DHCP server and detecting the attack**

1. Input: MF =(FD, FS,FA & I)
2. Output : For every strategy I € FA, I € FS, D € FD)
3. D = Each valid DHCP Server
4. IP= Internet protocol address
5. N= Number of mobile devices
6. FD= Frame DHCP server
7. If D € FD
8. IP→ N
9. endif
10. S= Number of available signatures in signature based Intrusion detection system (SIDS)
11. FS= Frame of signatures
12. FS ⊆ SIDS
13. I= Number & Types of attacks
14. For ( I=S; I ≤ FS; I++)
15. If I ⊆ FS
16. SIDS attack alert
17. endif
18. endfor
19. A= Number of signatures available in Anomaly based Intrusion detection system AIDS
20. FA= Frame of AIDS
21. FA ⊆ AIDS
22. For ( I =A; I ≤ FA ; I ++)
23. If I ⊆ FA
24. AIDS raises alert
25. If ( I ∉ FS & I ∉ FA)
26. No alert ( No attack)
27. endif
28. endif
29. endfor

### 4.1. Monitoring Process of Detecting Server ( DS)

The following rules collectively function to determine the anomalies.

**i.** **Pre-selected rules:** They help to detect those patterns, which are already stored in DS that apply to identify the inbound traffic.

ii. **Post-selected rules:** They refer to those patterns which are stored for detection of legitimate DHCP server that help to identify outbound traffic.

iii. **Parameterized rules:** They refer to many ingredients that help to set selected rules with unique value presented in the following:

a. **Validity ingredient:** It helps to detect attack if intruder modifies the contents of message.

b. **Time interval ingredient:** It helps to detect two types of attacks which are exhaustion attack and negligence attack. In exhaustion attack, the intruder increases message-sending rate. In negligence attack, intruder does not send the message. In addition, time interval for two consecutive messages is increased or decreased than allowed amount of time that gives sign of attack.

c. **Flooding ingredient:** It helps to identify attack on basis of noise and disturbance to be created in communication channel.

d. **Retransmission ingredient**: It helps to determine attack, if retransmission does not occur before specified timeout period.

e. **High transmission radio range ingredients**: It helps to determine SYN flood and wormhole attack, when intruder uses powerful radio sending message to further located node.

f. **Pattern replication ingredient:** It helps to detect attack when same patterns are repeated several time, it blocks the DOS attacks.

All of these ingredients collectively help to DS for detecting the attacks and figure 4 shows the process how to determine valid IP address and attack.

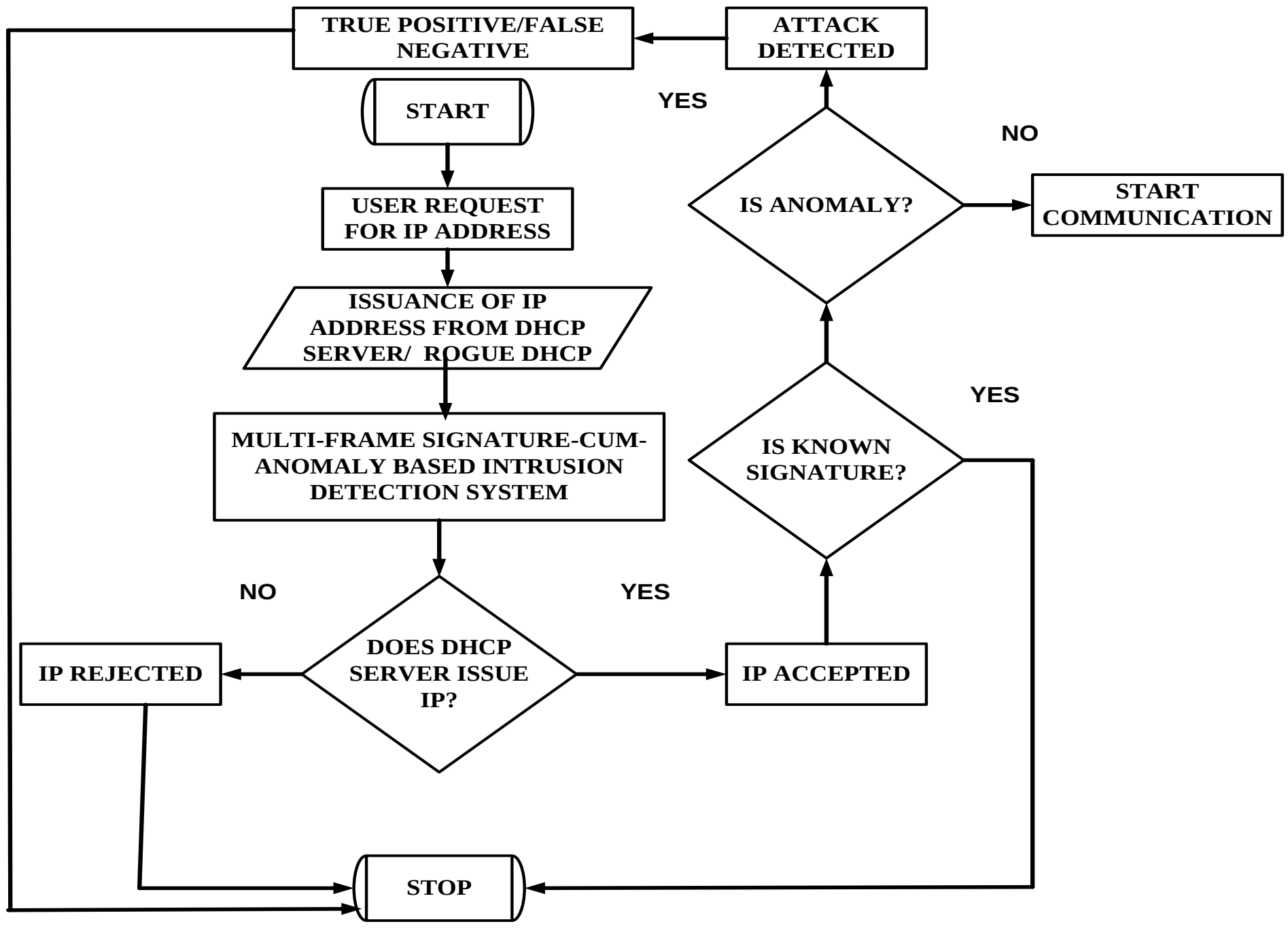

Figure 4. Detecting attack and issuance of safe IP

DS also controls the multi frame that comprises of central IDS and integrated with three layers that control the misuse detection.

### 4.2. Central IDS

The aim of central IDS is to control and store messages received from DS. It works as middleware for DS and other layers to send the verification request and receive alerts. The main function of central IDS is to update and manage the policy according to nature of attacks. If it needs any change in attack-detection that is employed on all the layers. The central IDS implements updated policy is shown in figure no 5.

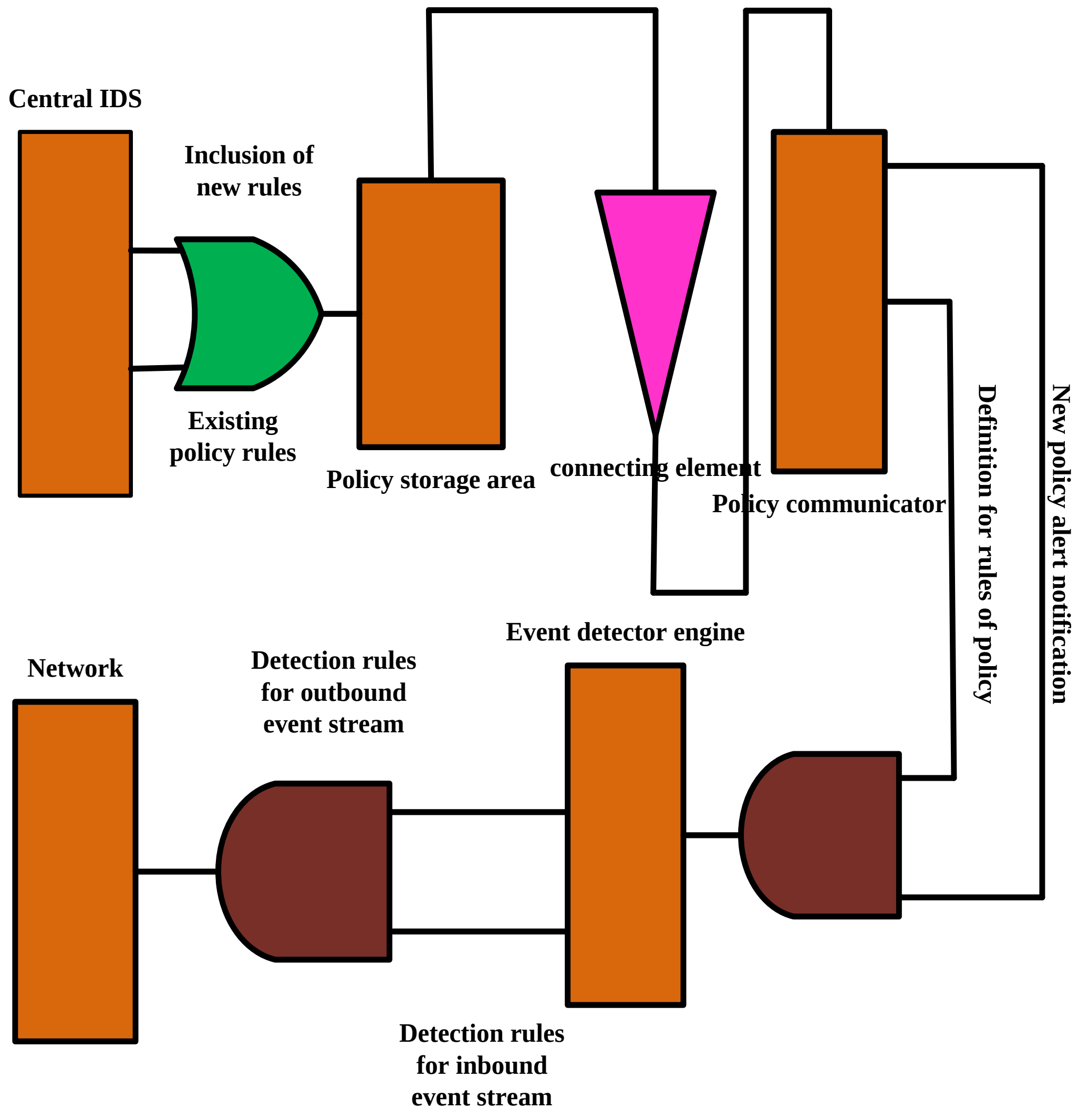

Figure 5. Policy of Central IDS for network

### 4.3. DHCP Verifier

DHCP verifier is top layer that distinguishes between rogue DHCP and original DHCP server. The signatures of original DHCP servers are stored at the DHCP verifier. It checks validity of DHCP server that issues IP address for client. On basis of stored signatures, DHCP server is identified whether it is rogue or original DHCP server. Top layer produces unique sign of alert for both DHCP rogue and original DHCP. Top layer receives parameters for verification from central IDS. DHCP verifier running on top layer is also responsible to return alert to central IDS.

### 4.4. Signature Based Detection Layer

Signature-based detection is middle layer that detects known threats. It compares signatures with observed events to determine possible attacks. Some known attacks are identified on basis of implemented security policy. For example, if telnet tries to use "root' username that is violating security policy of organization that is considered known attack. If operating system has 645 status code values that is sign of host's disabled auditing and refers as attack. If attachment is with file name "freepics.exe" that is alert of malware. Middle layer is effective for detection of known threats and using well-defined signature patterns of attack. The stored patterns are encoded in advance to match with network traffics to detect attack. This layer compares log entry with list of signatures by deploying string comparison operation. If signature based layer does not detect attack, anomaly based detection layer starts to process.

### 4.5. Anomaly Based Detection Layer

Lower layer is anomaly based detection that identifies unknown and DOS attacks. It works on pick-detect method. This method monitors inbound and outbound traffic Packets are evaluated, adaptive thresholds and mean values are set. It calculates the metrics and compares with thresholds [19]. On basis of comparisons, it detects various types of anomalies including false positive, false negative, true positive and true negative. If pick-detect methods determines true positive and false negative then it sends alert to Central IDS. The process of detecting anomalies is given in algorithm 2.

**Algorithm 2: Detecting the types of alerts with AIDS**

1. FA= Frame of Anomalies
2. FA ⊆ AIDS
3. Si = False Negative
4. Sj= True negative
5. Sk= True positive
6. Sk= False positive
7. 0 = don't match & 1= match
8. $Sijkl = 1/d \sum_{m=1}^{m} Sijkl$
9. Sij = { 0, if i & j
10. No false negative & true positive
11. Sij = { 1, if i & j
12. false negative & true positive
13. Alert of attack

14. Skl = { 0,   if   k & l [ do not match] & 1, if   k &   l [match]
15. Alert of true negative & false positive
16. No sign of attack
17. endif
18. endif
19. endif
20. endif

In addition to determine and calculate value of true positive and false negative; we apply algorithm 3 that helps to find attack and non-attack situation for TN and FN.

**Algorithm 3: Determine the sign of attack or non-sign of attack**

I. We select random odd prime number for TN and any even number for FN.
2. The value of FN must not be exceeded than TN.
3. Therefore, FN > 1 & FN< TN
4. Here, FN= {2, 4, 6, 8…} & TN= {3, 5, 7, 11, 13…}
5. Here sign of attack = ST, d = not exposed & b = exposed.
6. b and d has constant value 1.
7. Thus, ST = TN/ (TN+ d)/ FN (FN +b)
8. If value of ST > 1, it means there is no sign of attack, if the value of ST < 1 that is sign of attack.
9. endif

Assume FN = 2 & TN =3: BY applying the sign of attack formula:

ST = TN/ (TN+ d)/ FN (FN +b): Substitute the values in given formula.
ST = 3/ (3+ 1) / 2(2+1)
ST = 9/8
ST= 1.125
ST > 1

Here, ST > 1 means there is no sign of attack and we will be able to determine that is True negative (TN).

## 5. Simulation Setup

The previous sections have presented evidence of problems to be created by rogue DHCP server and including solutions to control problems. This section focuses on simulation setup and type of scenario. To validate the approach, the proposed solution has been implemented by using three methods: test bed simulation, discrete simulation in C++ and ns2 simulation.

We discuss only test bed simulation in this paper that provides real time results in controlled and live user environments. This kind of simulation gives complete understanding about behavior of several types of attacks. All operations associated with MSAIDS approach and other three existing approaches: Dynamic Multi-Layer Signature based IDS (DMSIDS), Ant Colony Optimization based IDS (ACOIDS) & Signature based IDS provide the recital idea. The parameters of test bed simulation are only given in table 1.

TABLE 1. Simulation parameters for test bed experiment.

| Name of parameters | Specification |
| --- | --- |
| MySQL database | MySQL 5.5 |
| Type of IDS | Rule based IDS |
| GD Library | gd 2.0.28 |
| Snort | V-2 |
| Apache web server | Apache http 2.0.64 Released |
| PHP | PHP 5.3.8 (Server side language) |
| ADODB | Release 5.12 (abstraction library for PHP and Python) |
| ACID | ACID PRO 7 |
| Stick | Stick beats detection tool used by hackers |
| Nikito | Nikito v.2.1.4 |
| IDS enabled system | Memory: 512 MB |
| | Operating system: Linux |
| | PCI network card: 10/100 Mbps |
| | CPU: P-III with 600 MHz |
| Attacker system | Memory: 1.5 GB |
| | Operating system: Linux |
| | PCI network card: 10/100/1000 Mbps |
| | CPU: AMD Geode LX running 2.4/5GHz |

In addition, most of operating systems do not provide the tracing facilities but regardless of problems, we would like to obtain result in standardized method by using different programs on different operating systems. MSAIDS has fully support of algorithms and data structure that discover potential attacks and perturb the intrusions before the attacks. The performance highly depends on robust tracing facility and algorithms, which help to identify the intrusion. The first step is to analyze performance of proposed algorithms. However, overall target is to obtain accurate statistical data in highly loaded network. Test bed simulation provides promising result. The mean value is calculated with help of following theorem.

**Theorem 1:**

Assume x = test bed simulation;
y= discrete simulation in C++ & ns2 simulation.
R is the proposed approach MASIDS.
Thus, Let f: [x, y] → R is the continuous function for closed interval [x, y]
Therefore
Let f: [x, y] → R is the differentiable continuous function for open interval (x, y)
Here x < y.
Hence z exists in (x, y)
Such that
f' (x) = f (x) – f (y)/( x- y)

## 6. Analysis of Result and Discussion

The training period of experiment covers four classes of attacks probe, DOS, U2R and R2L. All detected attacks are included in database during the training period in test bed simulation. Ns2 and discrete simulation in C++ provides tracing facility to collect accurate data. The MSAIDS scans all rules of snort and includes new rules explained in proposed section 3. The testing period targets one concise scenario. The scenario is simulated by using same parameters for all three existing approaches including our proposed MSAIDS. The attacks are generated by using stick, covering all types of signatures and anomaly based attacks.

The training period provides quite interesting results because frequently generated attacks are of different numbers. The maximum number of attacks pertains to R2L category. The more attacks are also counted on MSAIDS as compare with other three existing techniques are shown in table no.2. If attack is not generated then it is counted as normal traffic. The frequency of single and group characters is displayed when packets reach at the attacker machine. It is observed on the basis of output that different types of detected attacks are generated due to rogue DHCP server.

The DOS attacks are detected when packet does not reach at destination and received no acknowledgment. The sign of probe attack is addition of new data in existing amount of data bytes. U2R is the sign of maximum connection duration. R2L attacks are little bit complex to detect. We apply method comprises of service requested and duration of connection for network and attempts failed login for host. It shows that proposed approach does not restrict the generating ratio of packets. From other side, the proposed approaches provides highest capturing ratio. The statistical results show that MSAIDS will substantiate to medical field for diagnosing several disease and especially for heart. The major breakthrough of this research is to detect the true positive and false negative attacks because they are very hard to capture.

TABLE 2: Comparison of attacks on MSAIDS with other known techniques

| Type of generated attack | (MSAIDS) | (DMSIDS) | (ACOIDS) | Signature based IDS |
|---|---|---|---|---|
| DOS attacks | 34214 | 33542 | 33421 | 32741 |
| U2R attacks | 12454 | 11874 | 11845 | 11341 |
| R2L attacks | 34123 | 32123 | 31092 | 29984 |
| Probe attacks | 6214 | 8758 | 10181 | 4907 |

Due to these anomalies, confidentiality of any system is exploited and privacy of the user is exposed. The proposed method also captures the real worm attacks and all other looming attacks. The table 3 shows quantitative results for each scheme.

The major advantage of MSAIDS approach is to detect all types of anomalies and unknown threats efficiently. The systems are mostly infected due to new sort of malwares because they consume the processing resources of system. If resources of system are utilized by unnecessary programs then MCL is highly affected. In consequence, collaboration process is disrupted.

MSAIDS also detects the activity for any specific session. It creates specific alarm for each type of anomalies. Furthermore, deployed algorithms and new addition of rules in ordinary IDS improves the performance and restore the privacy of users.

TABLE 3: Showing quantitative data for proposed and existing schemes

| Parameters | (MSAIDS) | (DMSIDS) | (ACOIDS) | Signature based IDS |
|---|---|---|---|---|
| total no: of packet to be received | 236719 | 198678 | 201938 | 178109 |
| total no: of packet to be analyzed | 236456 | 192453 | 197842 | 170098 |
| total no: of attack to be generated | 87005 | 86293 | 86539 | 78973 |
| total no: of signature based attack to be generated | 42003 | 42287 | 42585 | 35098 |
| total no: of anomaly based attack to be generated<br>a. False positive<br>b. False negative<br>c. True positive<br>d. True negative | a. 9475<br>b. 12093<br>c. 18574<br>d. 4860 | a. 8574<br>b. 11987<br>c. 16943<br>d. 6502 | a. 8878<br>b. 12007<br>c. 16943<br>d. 6126 | a. 8458<br>b. 11289<br>c. 12455<br>d. 11673 |
| total number of anomaly based attack to be generated | 45002 | 44006 | 43954 | 43875 |
| total no: of attacks to be captured | 87002 | 84212 | 83434 | 36098 |
| anomaly based attacked captured<br>a. False positive<br>b. False negative<br>c. True positive<br>d. True negative | a. 9475<br>b. 12093<br>c. 18574<br>d. 4859 | a. 8324<br>b. 11456<br>c. 15678<br>d. 6324 | a. 8678<br>b. 11987<br>c. 16789<br>d. 6045 | a. 654<br>b. 245<br>c. 453<br>d. 535 |
| total number of anomaly based attacks to be captured | 45001 | 41782 | 43499 | 1887 |
| signature based attacks to be captured | 42001 | 42102 | 41091 | 34211 |

The proposed method also captures the real worm attacks and all other looming attacks. MSAIDS captures known attacks frequently. The figure 6 shows capturing capability of known attacks.

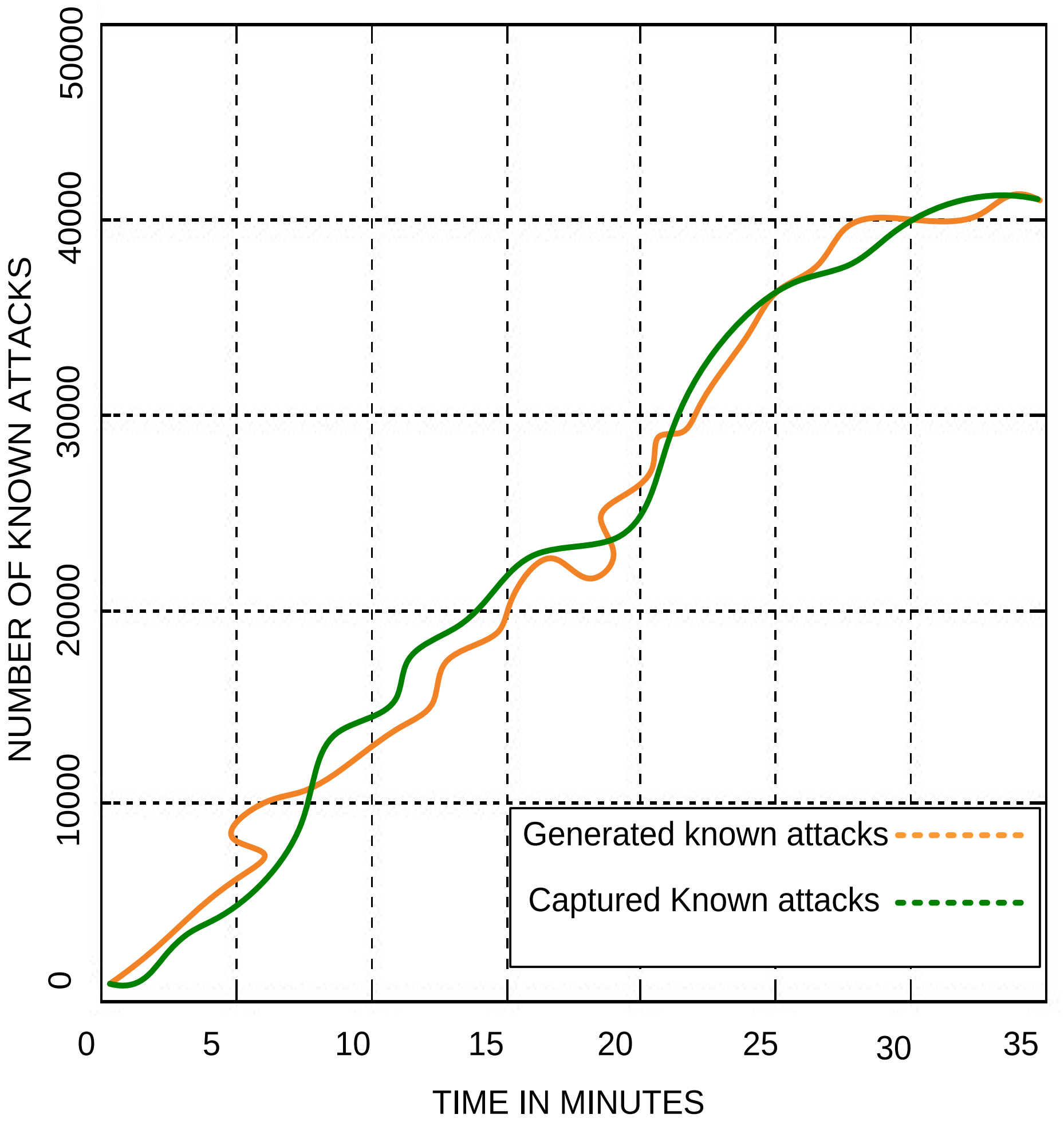

Figure 6: showing generating known attacks and capturing capability of MSAIDS

The major advantage of MSAIDS approach is also to detect all types of anomalies and unknown threats efficiently. The systems are mostly infected due to new sort of malwares because they consume the processing resources of system. If resources of system are utilized by unnecessary programs then communication is highly affected. In consequence, collaboration process is disrupted. MSAIDS also detects the activity for any specific session. It creates specific alarm for each type of anomalies. The beauty of this approach is high capturing capability of anomalies shown in figure 7.

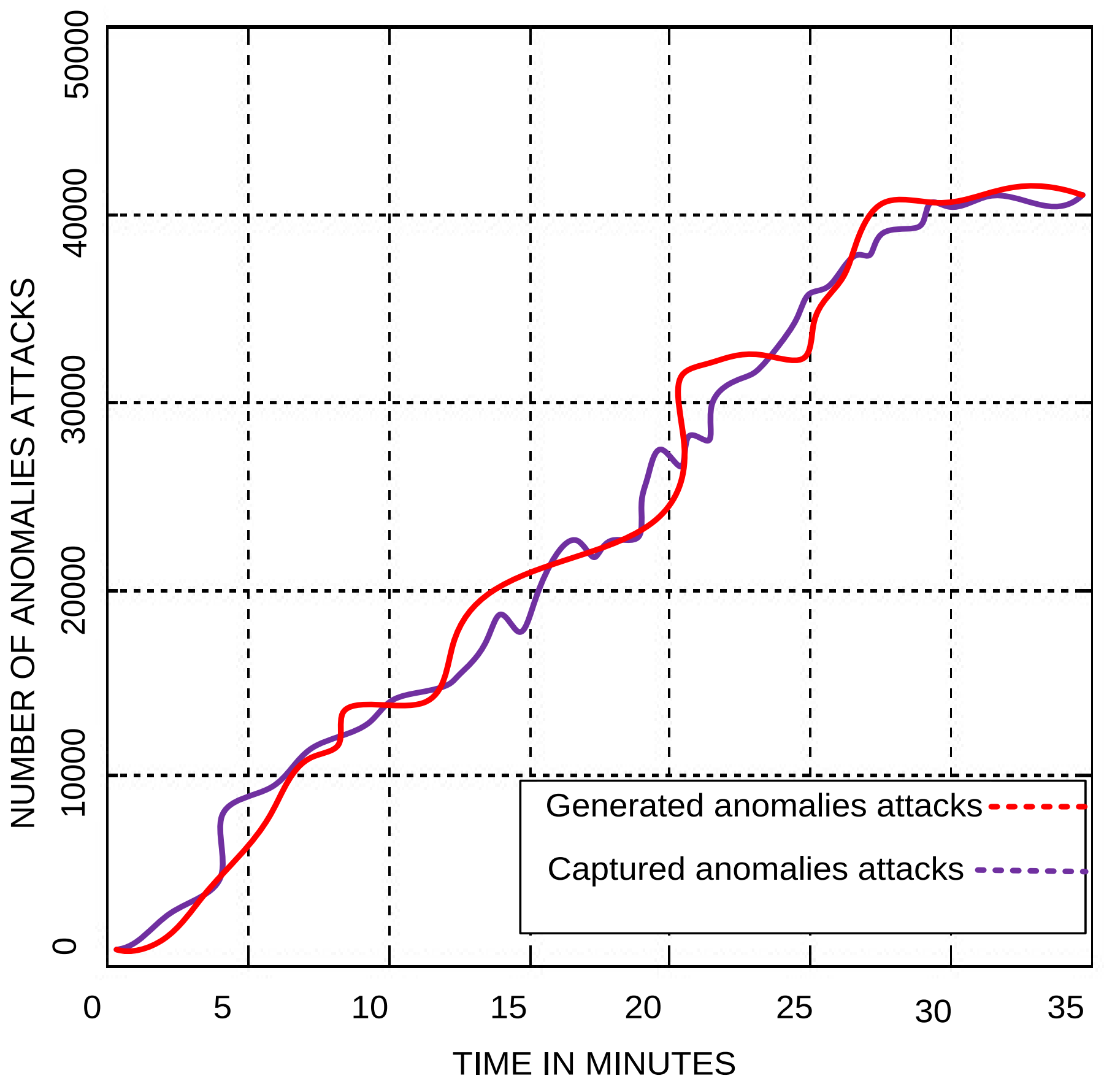

Figure 7: generation of Anomalies vs. capturing capability of MSAIDS

Furthermore, implemented algorithms and addition of some new rules in ordinary IDS improves the performance and restore the privacy of users. The implementation of MSAIDS is supported with sound architectural design that is robust and persistent when attack is detected. Statistical data shows 99.996% overall efficiency of MSAIDS shown in figure 8.

The efficiency of MSAIDS is calculated with following formula:

Here, overall efficiency = $E_a$;
Total generated signature based attacks = TSA;
Total anomaly based attacks =TAS;
Missed signature based attacks = MSA;
Missed anomaly based attacks = MAA & total generated attacks = TGA.

Thus, $E_a$ = (TSA + TAA) – MSA + MAA) * 100 / TGA

The proposed MSAIDS framework produces 2.269 to 49.11 higher capturing-rates than other existing techniques. The more interesting work of this research is detailed expression of all types of anomalies separately in form of false positive, false negative, true positive and true negative.

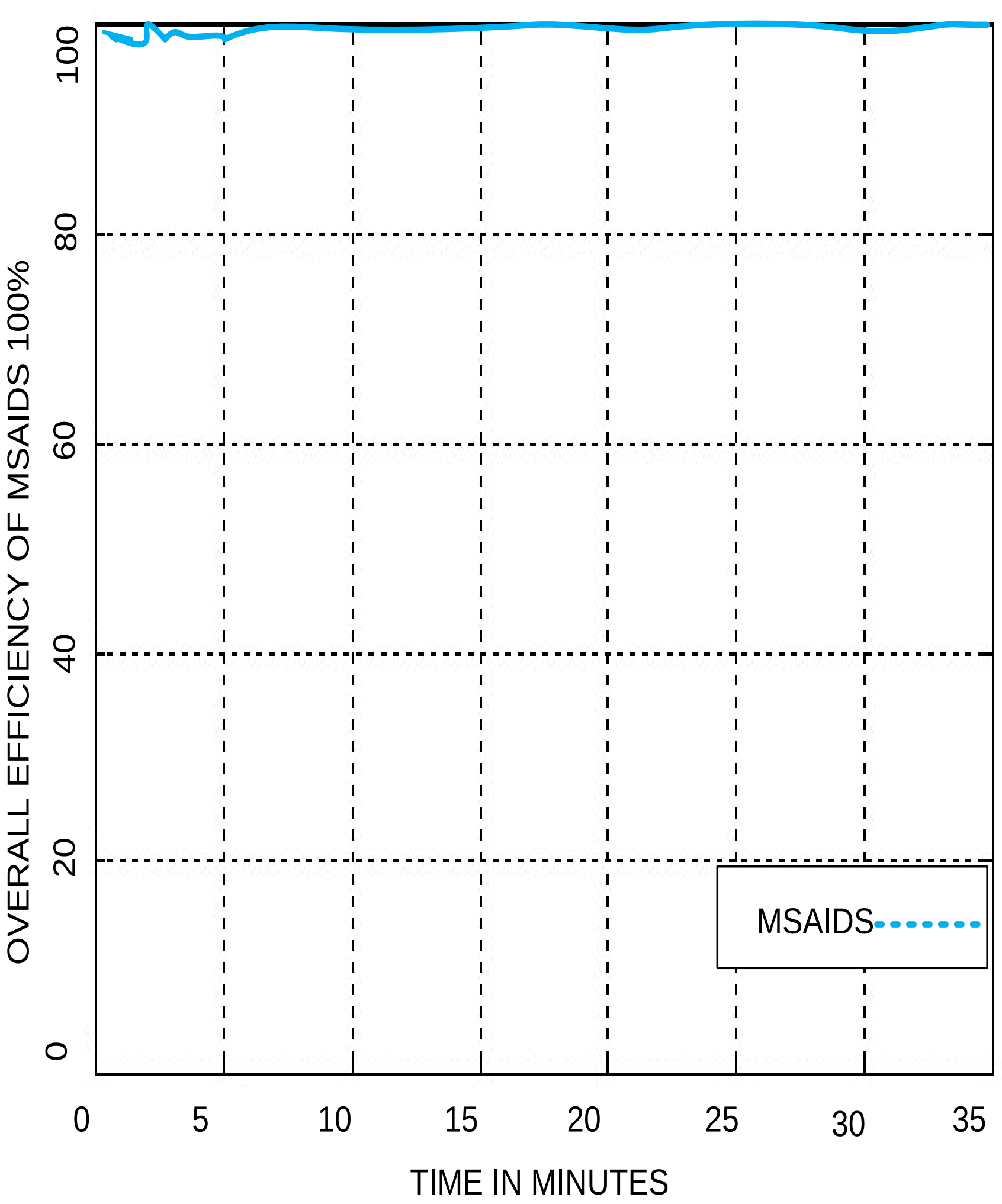

Figure 8. Comparison efficiency of all approaches

These parameters give the concrete idea to use the features for various types of applications in real environment. The figure 9 shows the capturing capacity of MSAIDS and other existing techniques with respect the time during the testing period.

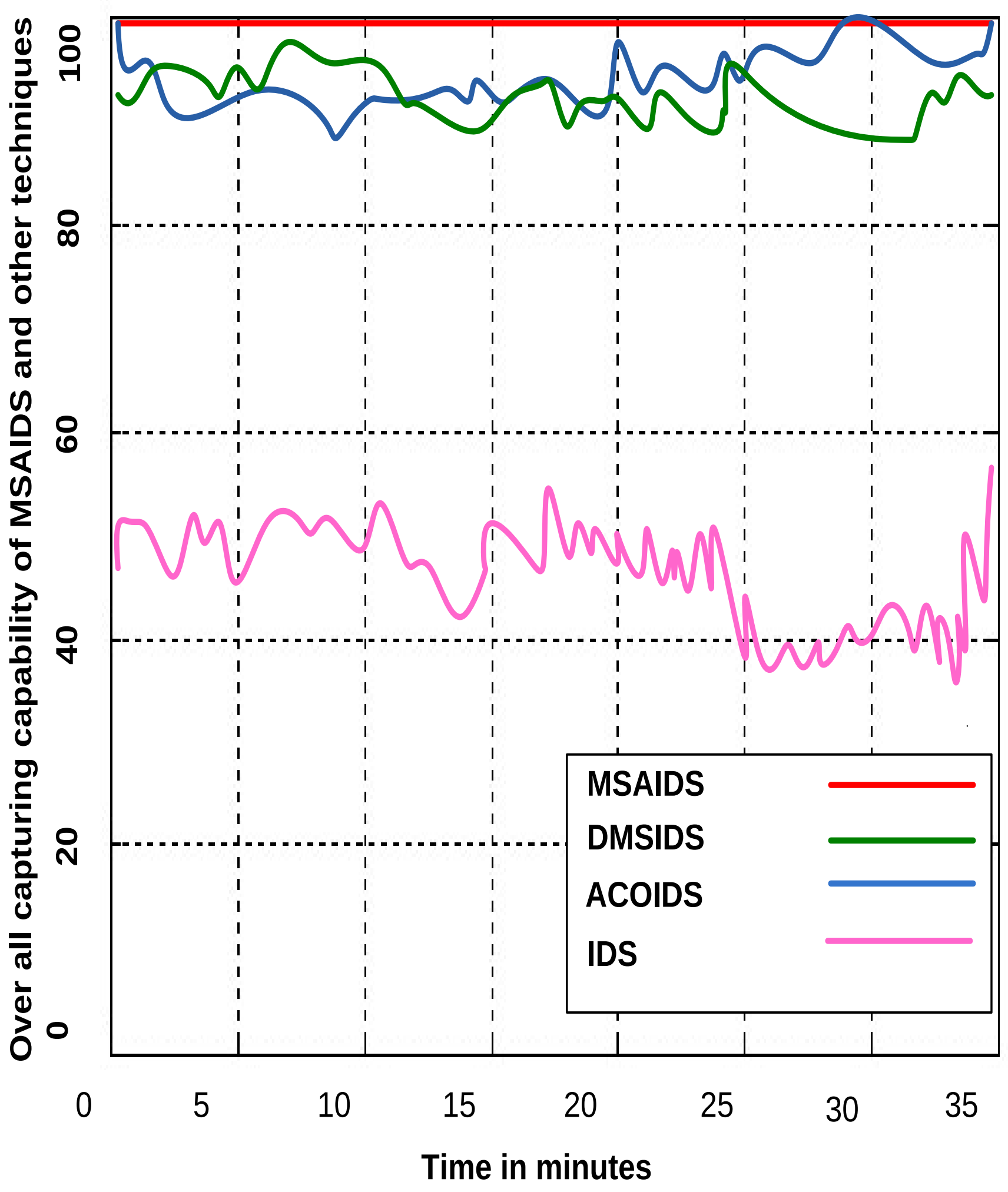

Figure 9: Overall efficiency of MSAIDS VS other existing techniques

## 7. Conclusion and Future Works

In this paper, multi-frame signature-cum anomaly-based intrusion detection systems (MSAIDS) is introduced. MSAIDS manages malicious activities of DHCP rogue server to restore privacy of users. The paper highlights malicious threats to be generated by DHCP rogue. The intruders use DHCP server to sniff traffic and finally deteriorate confidential information. The mechanism of current IDS does not have enough capability to control several types of malicious threats. Furthermore, several daunting and thrilling challenges in the arena of computer network security are impediment for secure communication. DHCP rogue is visibly very simple but crashes network as well as privacy of the users and even creates nastier attacks like Sniffing network traffic, masquerading attack, shutting down systems and DOS. The first is detailed explanation of attacks and how to resolve this issue. Second, we propose technique that is based on algorithms and addition of new rules in existing current IDS. These all of the components of proposal collectively handle the issues of DHCP rogue.

To validate the proposal, the technique is simulated by using test bed. Two different kinds of systems are used in test bed; as one is reserved for intruder and other one is for legitimate user. On basis of simulation, we obtain very interesting data, which show that MSAIDS improves capturing performance and controls attacks to be generated by DHCP rogue as compare with original IDS and other well known techniques. The findings demonstrate that MSAIDS has significantly reduced false alarms. Finally, we analyze overall efficiency of MSAIDS and existing techniques. In future, this technique will be deployed to measure the myocardits of heart.